\documentclass[twocolumn]{aastex631}
\received{October 16, 2023}
\revised{February 3, 2024}
\accepted{Februrary 20, 2024}

\begin{document}

\defcitealias{Biviano+17b}{B+17} 
\defcitealias{MBB13}{MBB13} 
\defcitealias{Poggianti+16}{P+16} 

\newcommand{\gtot}{\gamma_{{\rm tot}}}
\newcommand{\gdm}{\gamma_{{\rm DM}}}
\newcommand{\rtwo}{r_{200}}
\newcommand{\vtwo}{v_{200}}
\newcommand{\rs}{r_s}
\newcommand{\rv}{r_{\Delta}}
\newcommand{\vv}{v_{\Delta}}
\newcommand{\rl}{R_{\mathrm{lim}}}
\newcommand{\rfive}{r_{500}}
\newcommand{\rone}{r_{100}}
\newcommand{\rnu}{r_{\nu}}
\newcommand{\cnu}{c_{\nu}}
\newcommand{\rdue}{r_2}
\newcommand{\mv}{M_{\Delta}}
\newcommand{\mtwo}{M_{200}}
\newcommand{\mfive}{M_{500}}
\newcommand{\prmt}{M_{p}(0.2)}
\newcommand{\prmf}{M_{p}(0.5)}
\newcommand{\mpr}{M_p(<R)}
\newcommand{\ks}{\rm{km} \, \rm{s}^{-1}}
\newcommand{\msun}{M_{\odot}}
\newcommand{\lsun}{L_{\odot}}
\newcommand{\ctwo}{c_{200}}
\newcommand{\slos}{\sigma_{\rm{los}}}
\newcommand{\vrf}{v_{\mathrm{rf}}}
\newcommand{\mamp}{\texttt{MAMPOSSt}~}
\newcommand{\jei}{\texttt{JEI}~}
\newcommand{\jeii}{\texttt{JEI}}
\newcommand{\ab}[1]{\textcolor{blue}{\bf #1}}
\newcommand{\rev}[1]{\textcolor{red}{\bf #1}}

\title{The radial orbits of ram-pressure-stripped galaxies in clusters from the GASP survey}

\author[0000-0002-0857-0732]{Andrea Biviano}
\affiliation{INAF-Osservatorio Astronomico di Trieste, via G.B. Tiepolo 11, 34143 Trieste, Italy}
\affiliation{IFPU-Institute for Fundamental Physics of the Universe,
via Beirut 2,
34014 Trieste, Italy}

\author[0000-0001-8751-8360]{Bianca M. Poggianti}
\affiliation{ INAF–Osservatorio Astronomico di Padova, Vicolo Osservatorio 5, I-35122 Padova, Italy}

\author[0000-0003-2150-1130]{Yara Jaff\'e}
\affiliation{Departamento de Física, Universidad Técnica Federico Santa María, Avenida España 1680, Valparaíso, Chile}
\affiliation{Instituto de F\'isica y Astronom\'ia, Universidad de Valpara\'iso, Avda. Gran Breta\~na 1111, Valpara\'iso, Chile}

\author[0000-0002-4393-7798]{Ana C.C. Louren\c{c}o}
\affiliation{European Southern Observatory (ESO), Alonso de Cordova 3107, Santiago, Chile}
\affiliation{Instituto de F\'isica y Astronom\'ia, Universidad de Valpara\'iso, Avda. Gran Breta\~na 1111, Valpara\'iso, Chile}

\author[0000-0001-5654-7580]{Lorenzo Pizzuti}
\affiliation{Dipartimento di Fisica G. Occhialini, Universit\'a degli Studi di Milano Bicocca, Piazza della Scienza 3, I-20126 Milano, Italy}

\author[0000-0002-1688-482X]{Alessia Moretti}
\affiliation{ INAF–Osservatorio Astronomico di Padova, Vicolo Osservatorio 5, I-35122 Padova, Italy}

\author[0000-0003-0980-1499]{Benedetta Vulcani}
\affiliation{ INAF–Osservatorio Astronomico di Padova, Vicolo Osservatorio 5, I-35122 Padova, Italy}



\begin{abstract}
We analyse a sample of 244 ram-pressure-stripped candidate galaxy members within the virial radius of 62 nearby clusters, to determine their velocity anisotropy profile $\beta(r)$. We use previously determined mass profiles for the 62 clusters to build an {\em ensemble cluster} by stacking the 62 cluster samples in projected phase-space. We solve the Jeans equation for dynamical equilibrium by two methods, \mamp and the Jeans inversion technique, and determine $\beta(r)$ both in parametric form and non-parametrically. The two methods consistently indicate that the orbits of the ram-pressure-stripped candidates are increasingly radial with distance from the cluster center, from almost isotropic ($\beta \simeq 0$) at the center, to very radial at the virial radius ($\beta \simeq 0.7$). The orbits of cluster galaxies undergoing ram-pressure stripping are similar to those of spiral cluster galaxies, but more radially elongated at large radii.
\end{abstract}

\keywords{
Galaxy clusters (584)  --- Galaxy kinematics (602)
}


\section{Introduction} \label{s:intro}

%Galaxies evolve faster in clusters

%Evolution depends on their orbits

%Orbits have been determined for several classes of galaxies

%Jellyfish are mostly ram-pressure galaxies

%Ram-pressure is orbital dependent

%Orbits have been already estimated from phase-space consideration

%This is a direct measurement, without comparing with sims.

%Orbits of cluster galaxies:

%Solanes+01: HI galaxies
%References from my recent works

%GASP:

%Jaffe+18 (orbits phase-space)
%Bellhouse+19 (orbit in phase-space)
%Salinas+23 (tail direction)

%Numerical simulations of orbits affecting galaxy evolution:

%van den Bosch+99 (dyn frict)
%Bekki+02
%Tonnesen19
%Lokas20
%Smith+22a
%Smith+22b
%Wright+22

The galaxy population in clusters of galaxies has been known to be different from that in the field for a very long time \citep[][and reference therein]{Hubble36,Dressler80,Biviano00}. This difference concerns various galaxy properties, among which color, star-formation rate, and morphology are perhaps the most striking. It is attributed to an accelerated evolution of cluster galaxies as they interact among themselves, with the cluster gravitational potential, and with the hot intra-cluster medium \citep[see, e.g.,][]{Moran+07,Biviano11,Poggianti21}. Among the several physical processes that have been proposed to explain the accelerated evolution of galaxies in clusters, ram-pressure stripping is the one with the strongest observational support \citep[see, e.g.,][for a review]{BFS22}. Galaxies that orbit in clusters experience a drag force that removes some (or all) of their gas in the inter-stellar medium, a process first proposed by \citet{GG72} and since then studied in detail through analytical and numerical simulation studies \citep[e.g.,][]{FS80,AMB99,QMB00,Hester06,McCarthy+08,TB09,Arthur+19,SGB19,XDLHF20,TB21,Akerman+23}. When a tail of stripped gas and stars is visible in the ram-pressure-stripped galaxies, they are named 'jellyfish' galaxies \citep{ESE14}.

Simulations have shown that galaxies on different orbits across the cluster are affected in different ways by the different environmental effects at play. Elongated orbits are expected to characterize new arrivals in the cluster potential, as dynamical friction and the non-adiabatic growth of the host halo lead to shrinking the orbits of galaxies with time \citep{Gao+04b,OTH21}. 
%On the other hand, orbital decay is reduced by the interaction of the galaxy inter-stellar medium (ISM) with the ICM, that tend to circularize galaxy orbits \citep{SDLDB08}.
The efficiency of ram-pressure stripping in removing gas from infalling galaxies and thereby stimulating morphological transformation depends on galaxy orbits, being stronger for more radial orbits \citep{VCBD01,DRVHB10} and for galaxies that infall along large-scale structure filaments \citep{BMCBF13}. Since the ram-pressure strength varies across the orbit of a galaxy in a cluster, taking into account the galaxy orbits is important to better predict the efficiency of ram-pressure stripping (\citeauthor{Tonnesen19} \citeyear{Tonnesen19}, see also \citeauthor{BG06} \citeyear{BG06} for a review). 

The orbital dependence of the ram-pressure stripping efficiency is supported by observations. Based on the shape of the velocity dispersion profiles of different classes of cluster galaxies, \citet{Solanes+01} suggested that HI gas-deficient spirals in clusters move on more radial orbits than their gas-rich counterparts. \citet{Vulcani+17} analysed a sample of H$\alpha$ emitting galaxies in clusters, characterized by an offset between the peak of the H$\alpha$ emission and that of the UV-continuum, that they suggest is an indication of ram-pressure stripping. By comparison with numerical simulations, they conclude that the H$\alpha$ cluster galaxies belong to the quartile of satellites on most radial orbits. \citet{Jaffe+18} use the phase-space distribution of jellyfish galaxies in clusters of the WINGS (WIde-field Nearby Galaxy-cluster Survey) and OmegaWINGS \citep{Fasano+06,Moretti+14} surveys, to conclude that many of them formed via ram-pressure stripping while infalling into the clusters on highly radial orbits.

In this paper, we determine the orbits of the WINGS and OmegaWINGS ram-pressure-stripped (RPS) candidate galaxies by solving the Jeans equation for dynamical equilibrium \citep{BT87}. More precisely, we construct an {\em ensemble} cluster by stacking 62 clusters in projected phase-space, and derive the velocity anisotropy profile of the RPS candidates that we identify as cluster members,
\begin{equation}
\beta(r) \equiv 1 - (\sigma_{\theta}^2+\sigma_{\phi}^2)/(2 \, \sigma_r^2)
\end{equation}
where $\sigma_r$ is the radial component of the velocity dispersion tensor, and $\sigma_{\theta}$ and $\sigma_{\phi}$ are the two tangential components, that we assume to be the same (no rotation of the {\em ensemble} cluster).

The structure of this paper is the following. In Sect.~\ref{s:data} we describe our data set, the construction of the {\em ensemble} cluster by stacking, and the cluster membership assignment to the RPS candidates. In Sect.~\ref{s:method} we describe how we determine $\beta(r)$ by two methods. The results of our analysis are presented in Sect.~\ref{s:res}. In Sect.~\ref{s:disc} we summarize and discuss our results and provide our conclusions. Throughout this paper we adopt the following cosmological parameters: $\Omega_m=0.3, \Omega_{\Lambda}=0.7, H_0=70$ km~s$^{-1}$~Mpc$^{-1}$.

\section{The data set} \label{s:data}
We use the catalogs of RPS candidates from \citet[][P+16 hereafter]{Poggianti+16} and \citet{Vulcani+22}. \citetalias{Poggianti+16} identified 419
galaxies as RPS candidates, based on morphological evidence for gas stripping from optical images. Of these, 344 were located in cluster fields. Based on the importance of the stripping signature, galaxies were assigned to five classes, \texttt{JClass} increasing from 1 to 5 with increasing evidence of stripping. Visual examples of different \texttt{JClass} galaxies are given in \citetalias{Poggianti+16}.

\begin{figure}
\includegraphics[width=0.5\textwidth]{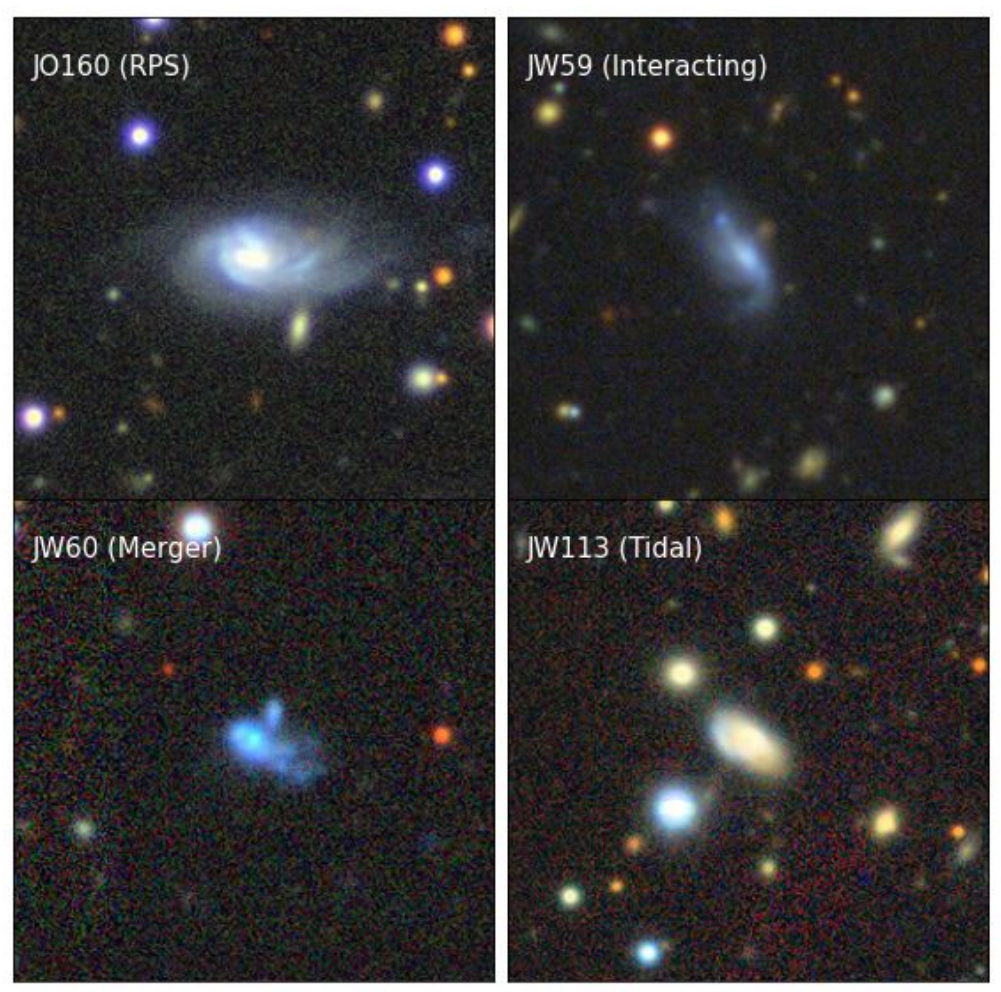}
\caption{Four examples of RPS candidates.  {\it Upper left panel: }\texttt{JClass=4} genuine RPS candidate. {\it Upper right panel:} \texttt{JClass=2} RPS candidate with morphological evidence of ongoing interactions. {\it Lower leftt panel:}\texttt{JClass=3} RPS candidate with morphological evidence of ongoing merger. {\it Lower right panel:} \texttt{JClass=2}
 RPS candidate with morphological evidence of tidal features. Composite grz optical images from Legacy Survey public data release DR10 \citep{Dey+19}).}
\label{f:rpstypes}
\end{figure}

Further visual inspection allowed \citetalias{Poggianti+16} to identify morphological features indicative of tidal stripping, mergers, and interactions with neighboring galaxies in
88 of the RPS candidates, $\sim 21$\% of the sample. Note that the evidence for tidal vs. ram-pressure stripping is not related to the \texttt{JClass}, since the latter only measure the evidence for stripping, which could be equally strong for the two different processes. In Fig.~\ref{f:rpstypes} we show four examples of RPS candidates, one for each of the interacting, merger, tidal, and genuine RPS categories.

The fraction of (possible) RPS contaminants obtained by the visual inspection, is confirmed by the study of Poggianti et al. (in prep.) who have conducted a spectroscopic survey with \texttt{MUSE@VLT} of a subset of the original sample of \citetalias{Poggianti+16}. They find out that 85\% of the RPS candidates in clusters are indeed galaxies undergoing ram-pressure stripping, as indicated by the presence of extra-planar ionized gas (i.e. H$\alpha$  emission) preferentially on one side of the disk while the disk stellar kinematics is undisturbed. The remaining 15\% of unconfirmed RPS candidates show disturbed stellar kinematics, in particular the absence of a regularly rotating stellar disk, probably caused by tidal effects, either mergers or a strong tidal interaction.

The fact that tidal effects are at work in some galaxies does not exclude that they are also subject to ram-pressure stripping. Indeed, one of these galaxies, JO134, is a clear example of the co-existence of two mechanisms, ram-pressure stripping and a minor merger event \citep[see][for further details]{Vulcani+21}. Another example is that of NGC~4654 in Virgo
\citep{Vollmer03}. To be conservative, we chose to remove the 88 possible RPS contaminants from our sample.

To this sample of 256 reliable RPS candidates we add the samples of 35 RPS candidates and of 143 unwinding spiral-arms galaxies (UG hereafter) identified by \citet{Vulcani+22}. Examples of UG are shown in Fig.~2 of \citet{Vulcani+22}. The UG are also considered to be RPS candidates, since ram-pressure stripping has been found to have an unwinding effect on the spiral arms \citep{Bellhouse+21}. Since there is no direct estimate of the confirmation rate of the unwinding RPS candidates, we discuss what is the effect of removing these UG from our sample in Sect.~\ref{ss:syst}.

The combined samples of \citetalias{Poggianti+16} and \citet{Vulcani+22}, after removing the 88 candidates with evidence of tidal or merging features, contain 434 RPS galaxies in the region of 68 clusters of the WINGS and OmegaWINGS surveys, at redshifts $0.04<z<0.07$; 350 of them have a measured redshift from the WINGS and OmegaWINGS data-sets \citep{Fasano+06,Moretti+14,Gullieuszik+15,Moretti+17}. 

\begin{figure}
\includegraphics[width=0.45\textwidth]{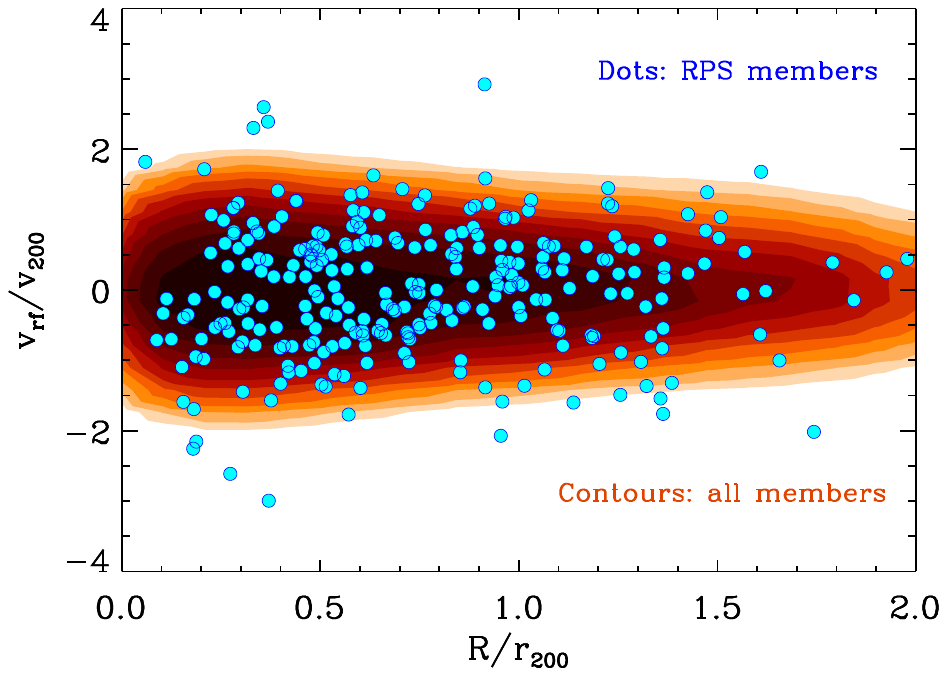}
\caption{Projected phase-space distribution of the {\it ensemble} cluster. The red-orange density contours, logarithmically spaced,  represent all cluster member galaxies, RPS galaxies excluded. The RPS galaxies selected as cluster members are indicated by blue dots.}
\label{f:rv}
\end{figure}

The 68 cluster centers are defined as the positions of the brightest cluster galaxies \citep[see][]{Fasano+10}. The mean cluster redshifts, $z_c$, and the virial radius, $\rtwo$\footnote{$\rv$ is the radius that encloses an average density $\Delta$ times the critical density at the halo redshift. $\mv$ is related to $\rv$ by  $\mv \equiv \Delta/2 \, \rm{H_z} \, \rv^3$/G, where $\rm{H_z}$ is the Hubble constant at the cluster redshift and G is the gravitational constant. The virial velocity is $\vv=10 \, \rm{H_z} \, \rv$.}  are taken from Table~B.1 in \citet[][B+17 hereafter]{Biviano+17b} for 47 clusters, and from \citet{Cava+17} for another 20 clusters not studied by \citetalias{Biviano+17b}. For A3164 we take the cluster redshift and $\rfive$ from \citet{Piffaretti+11}, and we convert $\rfive$ to $\rtwo$ by adopting a NFW profile \citep{NFW96} with a concentration $\ctwo=4$, typical of low-redshift massive clusters \citep[e.g.,][]{DeBoni+13}. We take the NFW profile scale radii $\rs$ from Table~B.1 in \citetalias{Biviano+17b} for 47 clusters, and we calculate it from $\rtwo$ using eq.~(11) in \citetalias{Biviano+17b} for the remaining 21 clusters. The mean redshift of these 68 clusters is 0.053 and the mean $\rtwo=1.6$ Mpc, corresponding to a mass $\log \mtwo/\msun=14.7$.

To define membership of the 350 RPS galaxies with $z$, we consider three methods.
The first method is based on a traditional $3 \, \sigma_v$ clipping, where $\sigma_v$ is the cluster line-of-sight (los hereafter) velocity dispersion \citep[listed in Table B.1 of][with an average fractional error of $\sim 6$\%]{Biviano+17a}. This method
was used by \citet{Paccagnella+17} on the WINGS and OmegaWINGS data-sets. For the other two methods, we follow the procedure described in \citet{Biviano+21}. First, we discard obvious interlopers by selecting galaxies with $\mid z-z_c \mid \leq 0.02$. On the $z$ distribution of these galaxies, we then apply the Kernel Mixture Model (KMM) algorithm \citep{McLB88,ABZ94}, to identify statistically significant secondary peaks, indicative of merging subclusters along the los and remove them from the sample. Finally we run two algorithms,  \texttt{Clean} \citep[][MBB13 hereafter]{MBB13} and \texttt{CLUMPS} \citep{Biviano+21} that select members based on the location of galaxies in projected phase-space, $R, \vrf$. $R$ is the projected radial distance from the cluster center, $\vrf$ is the rest-frame velocity $\vrf \equiv c \, (z-\overline{z})/(1+\overline{z})$, where $c$ is the speed of light, and  $\overline{z}$ is the mean cluster redshift defined in an iterative way on the selected cluster members. \texttt{Clean} is based on theoretically motivated models of the mass and velocity anisotropy distribution of clusters. The algorithm is statistically robust, but potentially subject to systematic bias, if the chosen models do not represent real clusters faithfully. \texttt{CLUMPS} searches for concentrations of galaxies in bins of projected phase-space, with no assumption about the internal dynamics of the cluster. However, since it is not based on a specific model, its results might depend quite sensitively on the choice of its parameters, which are calibrated to the size and radial extension of the cluster galaxy sample. 

We finally select a galaxy as a cluster member if it is considered a member by at least two of the three methods. It is in fact not advisable to rely on a single method only, as it might be subject to unknown systematic biases. On the other hand, by accepting only galaxies selected as members by all three methods we risk excluding radially infalling galaxies that are true cluster members mis-interpreted as interlopers \citep{Jaffe+18}.
Out of the 350 RPS galaxies with available $z$, we select 285 as members of 62 clusters - 6 clusters have no RPS member galaxies. To restrict our analysis to the cluster virial region (see Sect.~\ref{s:method}), we further select the 244 members located within $1.2 \, \rtwo$ (i.e., within $\sim r_{100}$) of their cluster center. 

The number of RPS member galaxies per cluster varies from 1 to 13, too small to allow determination of the RPS galaxy orbits in each individual cluster. We, therefore, need to stack the cluster data to build an {\em ensemble} cluster, following a well established procedure
\citep[e.g.][]{Carlberg+97-equil,BG03,MG04,KBM04,Rines+13,Biviano+16,Cava+17}. The stacking procedure is based on the quasi-homology of cluster mass profiles. These profiles are well represented by the NFW model. Since the NFW model concentration depends very little on the halo mass and redshift at the cluster mass scale \citep[e.g.][]{DeBoni+13,Ettori+19,Biviano+21}, the cluster mass profiles mostly depend on a single parameter, $\rtwo$. We build the {\em ensemble} cluster by normalizing each galaxy $R$ and $\vrf$ by its cluster $\rtwo$ and $\vtwo$, respectively. We display in Fig.~\ref{f:rv} the distributions of the RPS galaxies selected as cluster members, and of all the cluster members, RPS galaxies excluded, in the {\it ensemble} cluster. A two-dimensional Kolmogorov-Smirnov test \citep{Peacock83,FF87} gives a probability $<0.01$ that the projected phase-space distribution of RPS member galaxies is drawn from the same parent population as the distribution of all other cluster members.

%\begin{table*}
%\centering
%\caption{Properties of the cluster sample}
%\label{t:clusters}
%\begin{tabular}{cccccccc}
%  \hline
%  Id & RA & Dec & $z$ & $\rtwo$ & $\rs$ & $\vtwo$ & $\rl$ \\ 
%     & \multicolumn{2}{c}{[degrees (J2000)]} & & \multicolumn{2}{c}{[Mpc]} & [$\ks$] & [Mpc] \\
%\hline
%A1069 & 159.931 & -8.6867 & 0.06528 & $1.18 \pm 0.16$ & $0.57 \pm 0.85$ & $851 \pm 119$ & 1.84 \\
%\hline
%\end{tabular}
%\tablecomments{An asterisk next to the cluster name indicates that the value of $\rs$ is derived from the value of $\rtwo$ %through eq.~(11) in \citet{Biviano+17a}.}
%\end{table*}

\section{Methods of analysis}\label{s:method}
To determine the orbits of the RPS member galaxies of the {\em ensemble} cluster, we solve the Jeans equation for dynamical equilibrium in spherical symmetry \citep[see, e.g.,][]{BT87}. The assumption of spherical symmetry is justified by construction, as the {\em ensemble} cluster is built from 62 clusters irrespective of their orientation \citep[see][for a discussion of the spherical assumption in the case of an ensemble cluster]{vanderMarel+00}. On the other hand, the assumption of equilibrium is valid only if the number of galaxies in any given region of the cluster phase-space does not change, $\partial f(\mathbf{x,v,}t)/ \partial t=0$, where $f$ is the distribution function and $\mathbf{x,v}$ are the galaxy positions and velocities.
The equilibrium assumption is therefore invalid if the cluster is rapidly growing in mass, but this should not be the case for our low-redshift cluster sample. In fact, based on the theoretical model predictions by \citet{ZJMB09}, a cluster with the mean mass and at the mean redshift of our sample, is expected to have grown in mass by $\sim 10$~\% only, during the last $\sim 1$~Gyr, which is the cluster dynamical time \citep{Sarazin86}. Since accretion is most likely to occur inside-out, to minimize its effect on the cluster dynamical state, we restrict our dynamical analysis to the virial region, that we define as the inner $1.2 \, \rtwo$ region, corresponding to radii $\lesssim r_{100}$. 
To check the equilibrium assumption we compare the results of our dynamical analysis on the 62 cluster sample, with the results we obtain on a subsample of 53 clusters from which we remove nine merging and post-merger clusters identified by \citet[][see Sect.~\ref{ss:syst}]{Lourenco+23}. 

To solve the Jeans equation we consider two methods. The first one is the \mamp method \citepalias{MBB13}. Given models for the mass profile, $M(r)$, and the velocity anisotropy profile, $\beta(r)$, and assuming Gaussianity of the velocity distribution of the galaxies in 3D, \mamp evaluates the probability of finding a cluster galaxy at its observed position in projected phase-space. By maximizing the product of all cluster member probabilities, \mamp constrains the parameters of the $M(r)$ and $\beta(r)$ models. \mamp has successfully been tested on simulated halos from cosmological simulations that included both dynamically relaxed and unrelaxed halos (\citetalias{MBB13}, \citeauthor{Tagliaferro+21} \citeyear{Tagliaferro+21}), and it has already been applied to several data sets \citep[e.g.][]{Biviano+13,Guennou+14,MBM14,Verdugo+16,Biviano+17a,Pizzuti+17,Mamon+19,Sartoris+20}. 

The second method we consider is the inversion of the Jeans equation \citep[\jei hereafter,][]{BM82,SSS90,DM92}. \jei requires knowledge of $M(r)$, but unlike \texttt{MAMPOSSt} it does not require fixing a model for $\beta(r)$. We follow the \jei method of \citet{Biviano+21}. This method has already been applied to several data-sets \citep[e.g.][]{BK04,Biviano+13,Annunziatella+16,Biviano+16,Zarattini+21}

The observable on which \mamp and \jei base their inference is the projected phase-space distribution of cluster members. In particular, both \mamp and \jei use the radial number density profile to describe the spatial distribution of cluster members, but while \mamp uses the full velocity distribution of cluster galaxies as a function of the galaxy radial distances, \jei only uses the velocity dispersion profile of cluster galaxies. It is the use of the full velocity distribution that allows \mamp to break the so-called mass-anisotropy degeneracy intrinsic to the Jeans equation, and to predict both $M(r)$ and $\beta(r)$ at the same time, unlike \jei that requires knowledge of $M(r)$.

Another difference between \mamp and \jei is in the treatment of the data. \mamp fits a model to the galaxy number density profile, and uses the individual galaxy velocities (no binning), while \jei performs a smoothing of the galaxy number density and velocity dispersion profiles \citep[we use the \texttt{LOWESS} smoothing technique, see, e.g.,][]{Gebhardt+94}. We extrapolate the smoothed profiles to large radii (30 Mpc) following \citet{Biviano+13}, to allow solving the equations that contain integrals up to infinity.

The uncertainties in the best-fit parameters of \mamp are obtained by a Monte Carlo Markov Chain (\texttt{MCMC}) analysis, as in \citet{Mamon+19}. The uncertainties in the (non parametric) $\beta(r)$ obtained by \jei are estimated by 300 bootstrap resamplings of the original data set \citep{ET86}.

\section{Results}\label{s:res}
\subsection{The number density profile}\label{ss:densprof}
To determine the projected number density profile ($N(R)$ in the following) of RPS member galaxies we need to take into account possible sources of incompleteness. While the original combined photometric sample of RPS candidates of \citet{Poggianti+16} and \citet{Vulcani+22} is complete down to the limiting magnitude of the survey, the spectroscopic sample is not. The incompleteness of the spectroscopic sample can affect the spatial distribution of the RPS member galaxies in such a way that their derived $N(R)$ is not representative of the parent sample. On the other hand, the velocity distribution is not affected by incompleteness, since the observational selection does not operate in redshift space within the narrow redshift range spanned by each cluster. 

\begin{figure}
\includegraphics[width=0.45\textwidth]{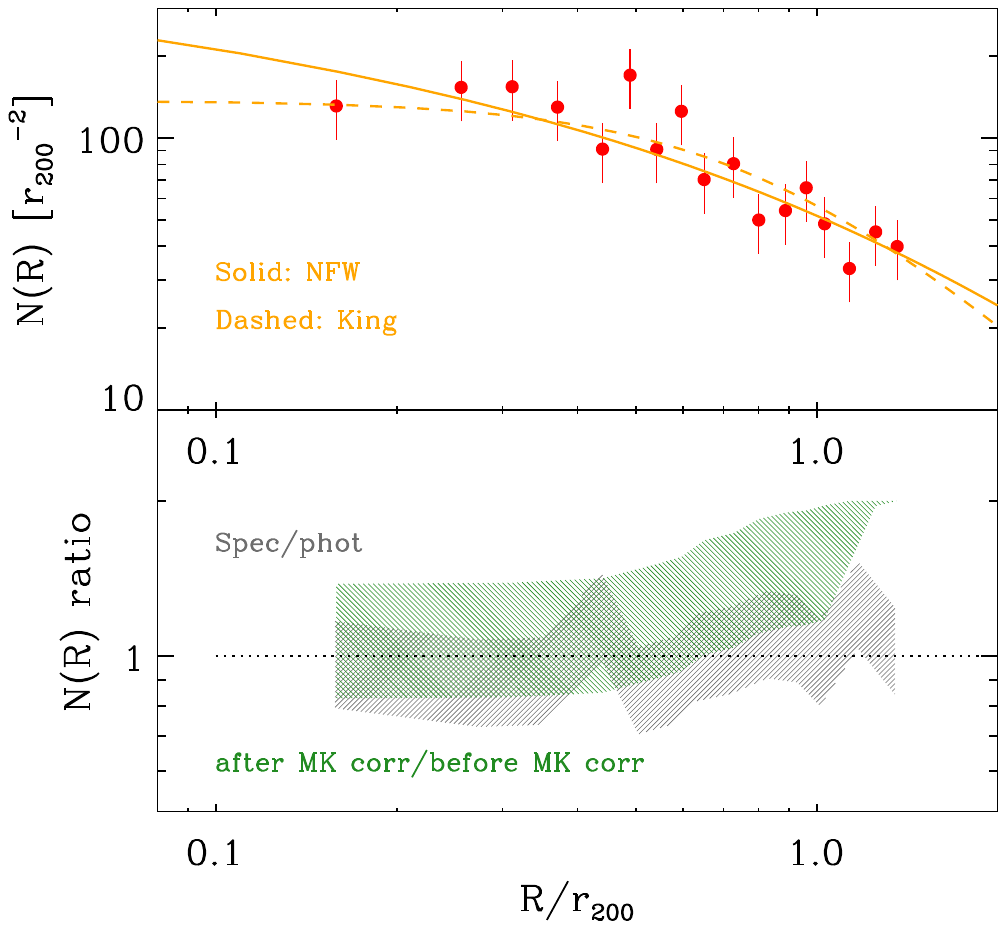}
\caption{{\it Upper panel:} Red dots: number density profile of the spectroscopic sample of RPS member galaxies in the {\em ensemble} cluster, corrected for the  \citet{MK89} incompleteness. Orange solid (dashed) line: best fit by a projected NFW (respectively, King) profile. {\it Lower panel:} grey shading: $\pm 1 \sigma$ contour of the ratio of the number density profiles of the spectroscopic sample of 350 RPS galaxies and of the photometric sample of 433 RPS galaxies. The ratio has been multiplied by the factor 433/350 to account for the different number of galaxies in the two samples.
Green shading: $\pm 1 \sigma$ contour of the ratio of the number density profiles of the spectroscopic sample of RPS member galaxies after and before the correction for the \citet{MK89} incompleteness.}
\label{f:nr}
\end{figure}

To assess possible biases in the spectroscopic selection, we compare the $N(R)$ of the 433 RPS galaxies from the photometric sample, with that of the 350 RPS galaxies from the spectroscopic sample. We find that the two $N(R)$ are almost identical, except for a different normalization reflecting the different number of galaxies (see grey shading in the bottom panel of Fig.~\ref{f:nr}). Since the normalization of the $N(R)$ is irrelevant for the Jeans analysis, we conclude that we do not need to apply any correction for the spectroscopic incompleteness.

Another source of incompleteness comes from the fact that in the {\it ensemble} cluster, not all clusters do extend to the same limiting radius. If not properly accounted for, this incompleteness effect would produce an artificial steepening of the {\it ensemble} cluster $N(R)$ at larger radii. The correction for this effect assumes that in the radial range where a cluster does not have data, its contribution can be ‘invented’ from the clusters that do have data, as first described by \citet{MK89}. Following \citet{KBM04}, we invent data only from measured data, and not from data that are themselves (partly) invented. As we show in Fig.~\ref{f:nr} (bottom panel, green shading), this incompleteness correction becomes significant at $R/\rtwo \gtrsim 0.7$.  In Fig.~\ref{f:nr} (upper panel) we show the $N(R)$ of the sample of RPS members, after correction for the \citet{MK89} incompleteness. We also show two best-fits, one with a projected NFW model \citep[][solid line]{Bartelmann96}, with a scale radius $r_g=1.8_{-0.4}^{+0.6}$ Mpc, and another with a \citet{King62} model (dashed line) with a core radius $r_{g,K}=0.83 \pm 0.05$ Mpc. Both fits are acceptable ($\chi^2=12.8$ and $13.0$, respectively, for 16 degrees of freedom) and we adopt the best of the two (NFW) in the \mamp analysis, after Abel de-projection in spherical symmetry to the 3D number density profile $\nu(r)$ \citep[e.g.][]{BT87}.

\subsection{The \mamp solution for $\beta(r)$}\label{ss:mamp}
We run \mamp using the NFW number density profile $\nu(r)$ described in Sect.~\ref{ss:densprof}, and
the $M(r)$ described by an NFW model with $\rtwo$ and $\rs$ parameters derived from the weighted averages of the $\rtwo$ and $\rs$ values of the 62 clusters that compose our {\em ensemble} cluster, weighting these values by the inverse of their uncertainties. These average values are $\langle \rtwo \rangle=1.63 \pm 0.03$ Mpc, and $\langle \rs \rangle=0.46 \pm 0.04$ Mpc, corresponding to an average concentration $\langle \ctwo \rangle=3.5$. We adopt the following, rather generic, model for  $\beta(r)$ \citep{Tiret+07}, 
\begin{equation}
    \beta = \beta_0 +\beta_{\infty} r/(r+r_{\beta}), \label{e:tiret}
\end{equation}
where we force $r_{\beta}=\rs$ as indicated by numerical simulations \citep{MBM10}, to reduce the number of free parameters in the \mamp analysis.

\begin{table}
\begin{center}
\caption{The \mamp parameters}
\label{t:mamp}
\begin{tabular}{cc}
\hline
Parameter & range \\
\hline
$\rtwo$ & 1.60 - 1.66 [Mpc]  \\
$\rs$ & 0.42 - 0.50 [Mpc] \\
$\rnu$ & 1.45 - 2.44 [Mpc] \\
$(\sigma_r/\sigma_{\theta})_0$ & 0.5 - 10.0 \\
$(\sigma_r/\sigma_{\theta})_{\infty}$ & 0.5 - 10.0 \\
\hline
\end{tabular}
\end{center}
\tablecomments{For each parameter we adopt a flat prior within the indicated range. The ranges for the $\rtwo$ and $\rs$ parameters are fixed to the $\pm 1 \sigma$ interval around the values obtained by the weighted mean of the 62 individual cluster values that compose the {\em ensemble} cluster. The range for the $\rnu$ parameter is fixed to the $\pm 1 \sigma$ interval around the best-fit value obtained by the maximum likelihood fit to the {\em ensemble} cluster $N(R)$ (see Sect.~\ref{ss:densprof}).} 
\end{table}

We determine the marginal distributions of the free \mamp parameters using the  \texttt{MCMC} technique, by sampling of 100,000 points in the parameter space.
We adopt flat priors for all parameters, but restrict the allowed range of the $M(r)$ and $\nu(r)$ parameters to their previously determined $\pm 1 \sigma$ intervals, while we allow a wide range for the $\beta_0$ and $\beta_{\infty}$ parameters, $\beta \in [-3,1[$.
For purely computing purposes, in our \texttt{MCMC} analysis we prefer to use the related parameters $\sigma_r/\sigma_{\theta}$ at $r=0$ and at $r \rightarrow \infty$, instead of $\beta_0, \beta_{\infty}$. In Table~\ref{t:mamp} we list the \mamp parameters and their ranges.

%\begin{table}
%\begin{center}
%\caption{Results of the \mamp analysis}
%\label{t:mamp}
%\begin{tabular}{crrr}
%\hline
%parameter & median & 68\% c.l. & 95\% c.l. \\
%\hline
%$(\sigma_r/\sigma_{\theta})_0$ & & & \\
%$(\sigma_r/\sigma_{\theta})_{\infty}$ & & & \\
%\end{tabular}
%\end{center}
%\tablecomments{Median values and 68\% and 95\% c.l.
%of the two velocity anisotropy  parameters of the $\beta(r)$ model eq.~\ref{e:tiret}, %marginalized using 100,000 MCMC.}
%\end{table}

\begin{figure}[ht]
\includegraphics[width=0.45\textwidth]{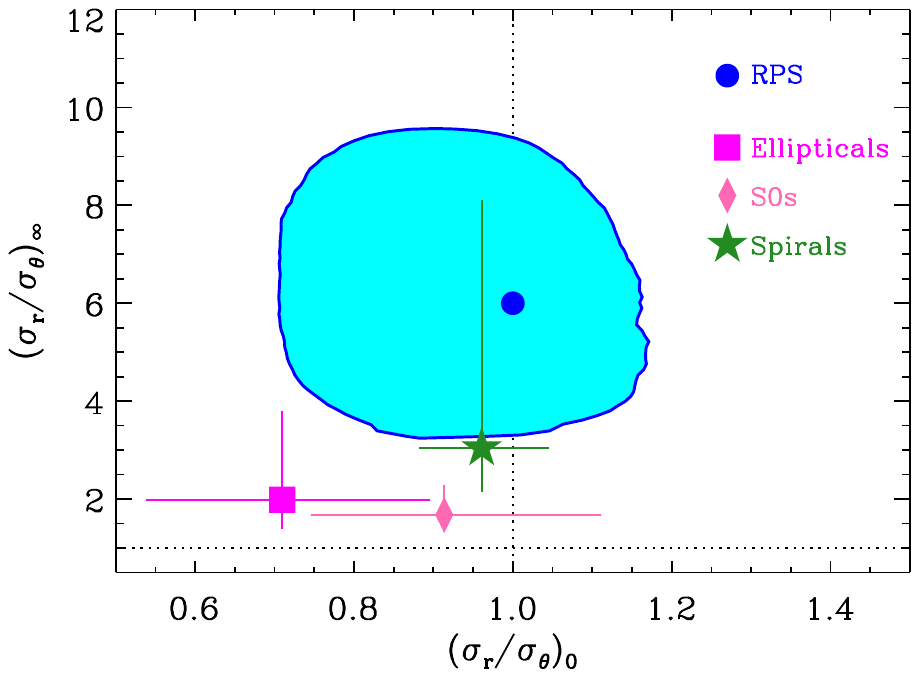}
\caption{Median values (blue dot) and 68\% confidence interval (cyan region)
of the two velocity anisotropy  parameters of the $\beta(r)$ model eq.~(\ref{e:tiret}), 
from the marginal distribution obtained with the \texttt{MCMC} \mamp analysis. The dotted lines indicate the isotropic values. The symbols with 1 $\sigma$ error bars represent the values for ellipticals (magenta square), S0s (pink diamond), and spirals (green star) in WINGS clusters, from Table~3, model~1 of \citet{Mamon+19}.}
\label{f:betap}
\end{figure}

\begin{figure}
\includegraphics[width=0.45\textwidth]{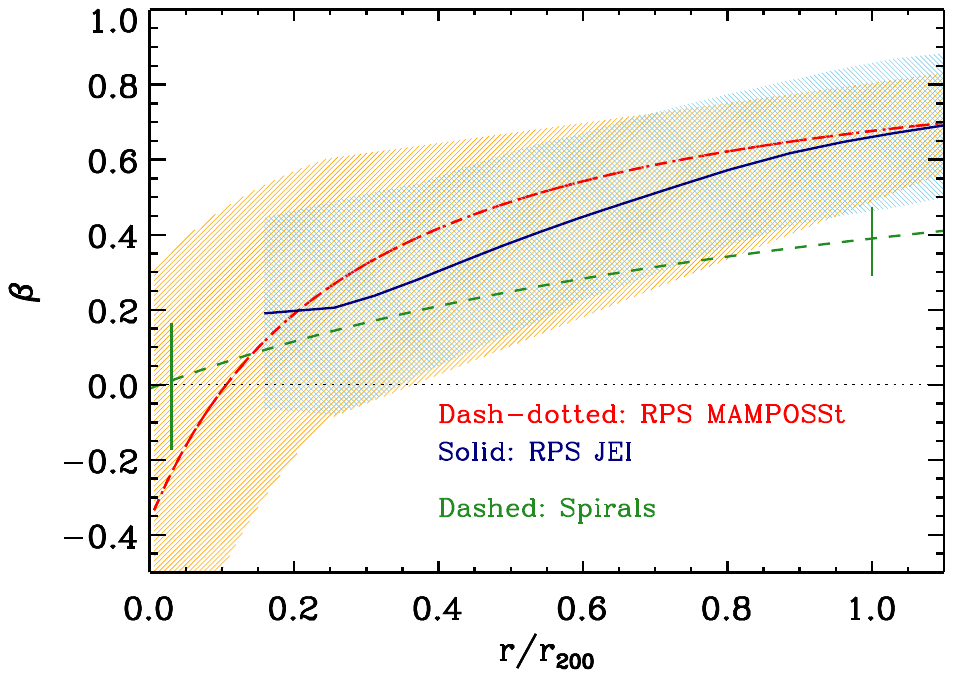}
\caption{Median $\beta(r)$ (red dash-dotted line) and its 68 \% confidence region (orange shading), as obtained from the \mamp analysis with \texttt{MCMC} sampling. Median $\beta(r)$ (blue solid line) and its 68\% confidence region (cyan shading), as obtained from the \jei analysis with bootstrap resamplings. For comparison we show the $\beta(r)$ for spirals in WINGS clusters (green dashed line) and its uncertainties at $0.03 \, \rtwo$ and $\rtwo$ (vertical grey segments), from Table~3, model~1 of \citet{Mamon+19}.
The dotted line represents orbital isotropy.}
\label{f:betaprof}
\end{figure}

\begin{figure}
\includegraphics[width=0.45\textwidth]{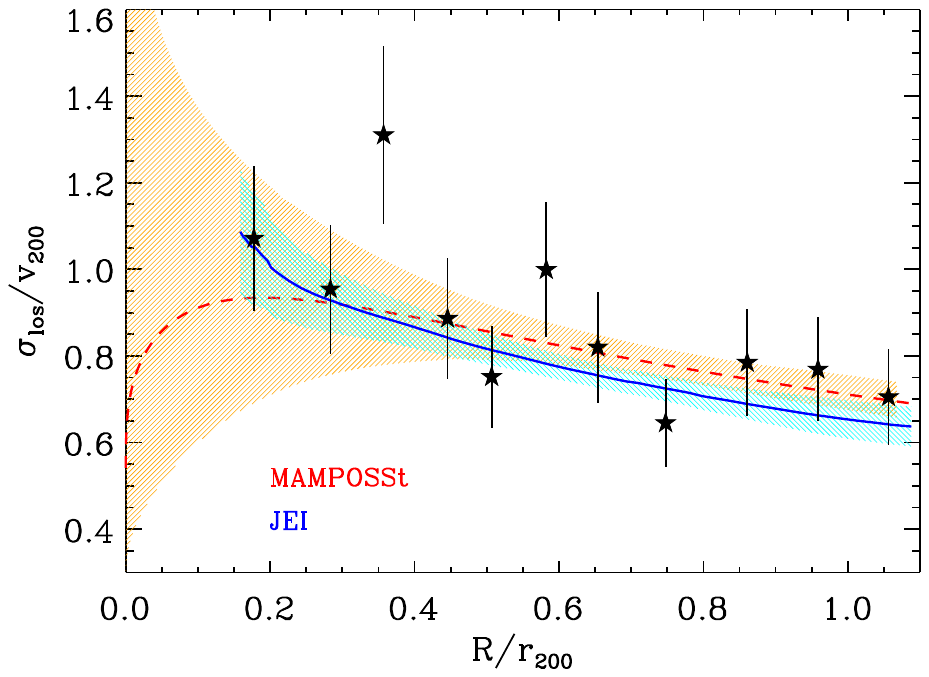}
\caption{Velocity dispersion profile of the RPS candidates (dots with 1 $\sigma$ error bars) compared to the predicted velocity dispersion profiles and their 68 \% confidence levels
of the \mamp (red dashed line and orange shading) and \jei (blue line and cyan shading) analyses.}
\label{f:vdp}
\end{figure}

The median value and 68\% confidence interval (in brackets) of the two velocity anisotropy parameters, from the marginal distribution obtained with the \texttt{MCMC} \mamp analysis, are $(\sigma_r/\sigma_{\theta})_0=1.0 \, [0.8,1.3]$ and $(\sigma_r/\sigma_{\theta})_{\infty}=6 \, [4,9]$, corresponding to $\beta_0=-0.05 \, [-0.75,0.37]$ and $\beta_{\infty}=0.97 \, [0.94,0.98]$. These results are displayed in Fig.~\ref{f:betap}. In Fig.~\ref{f:betaprof} we show the corresponding median $\beta(r)$ and its 68\% confidence region (red dashed line and orange shading). To check if the \mamp result is a good fit to the data, we project the \mamp result on the los velocity dispersion profile, $\sigma_{\mathrm{los}}(R)$ via \citep{BT87,vanderMarel94}
\begin{eqnarray}
\nu(r) \sigma_r^2(r) = \int_{r}^{\infty} \frac{G \nu M}{x^2} \exp \Bigl[ 2 \int_b^x \frac{\beta(t)}{t} \, dt \Bigr] \, dx, \\
%\end{equation}
%and
%\begin{equation}
    N(R) \sigma_{\mathrm{los}}(R) = 2 \int_R^{\infty} \frac{r \nu \sigma_r^2}{(r^2-R^2)^{1/2}} \, dr.
\end{eqnarray}
The \mamp $\sigma_{\mathrm{los}}(R)$ is compared to the observed one in Fig.~\ref{f:vdp} (red dashed line and orange shading and black dots, respectively). The $\chi^2$ goodness of fit test does not reject the null hypothesis that the data follow the model ($\chi^2=16.4$ for 12 degrees of freedom, corresponding to an 82~\% probability).

\subsection{The \jei solution for $\beta(r)$}\label{ss:jei}
We adopt the same NFW $M(r)$ used for the \mamp analysis (see Sect.~\ref{ss:mamp}). The $N(R)$ and $\sigma_{\mathrm{los}}(R)$ are \texttt{LOWESS}-smoothed version of the data (dots with error bars) shown in Figs.~\ref{f:nr} and \ref{f:vdp}, respectively. We bootstrap the data 300 times to evaluate the uncertainties. The results are shown in Fig.~\ref{f:betaprof} (blue line and cyan shading). Using the same procedure adopted for the \mamp solution, we project the \jei solution to compare the predicted and observed $\sigma_{\mathrm{los}}(R)$ (see Fig.~\ref{f:vdp}; the \jei prediction is the blue line with cyan shading indicating the 68\% confidence levels). The smaller uncertainties in the \jei solution result in a slightly higher $\chi^2$ value (19.4 for 12 degrees of freedom, corresponding to a 92~\% probability) with respect to the one of the \mamp result, but also in this case the null hypothesis is not rejected, the model is a good enough fit to the data.

\subsection{Systematics}\label{ss:syst}
We here consider three possible systematic effects affecting our result. One is the inclusion of the UG in our sample of RPS candidates. As discussed in Sect.~\ref{s:data}, we have no estimate of which fraction of the UG are indeed RPS galaxies. The second effect we investigate is the inclusion of merging and post-merger clusters in our sample. As discussed in Sect.~\ref{s:method} this could invalidate our analysis based on the Jeans equation for a system in dynamical equilibrium. 
Finally, we consider the impace of removing low \texttt{JClass} RPS canidates from our sample. By removing the UG we are left with 134 RPS spectroscopic members within $1.2 \, \rtwo$ (the no-UG subsample). 
By removing the 9 merging and post-merger clusters identified by \citet{Lourenco+23} we are left with 206 RPS spectroscopic members within $1.2 \, \rtwo$ of 53 clusters (the no-merging subsample). By removing the \texttt{JClass}=1 and 2 RPS candidates only 68 cluster members remain in our sample (the high-\texttt{JClass} sample).

\begin{figure}
\includegraphics[width=0.45\textwidth]{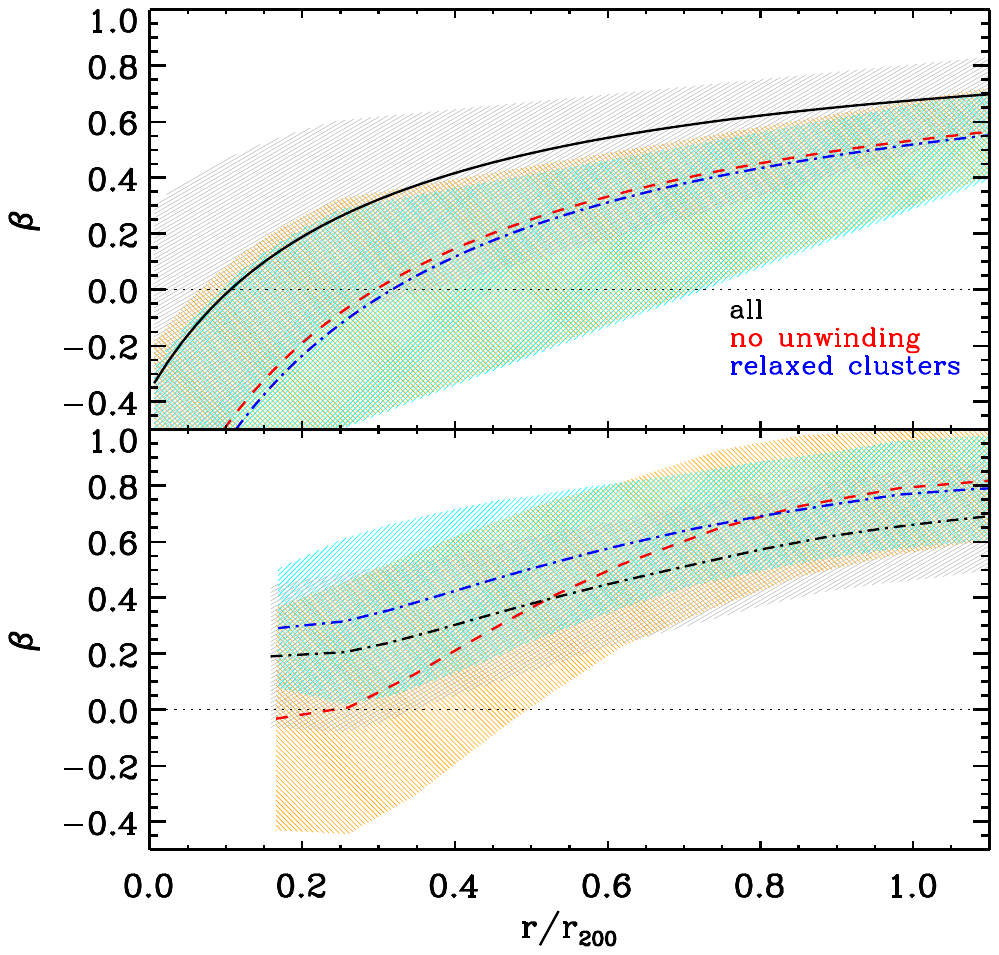}
\caption{Median $\beta(r)$ and its 68\% confidence region for the whole sample (solid black line and grey shading), the no-UG subsample (dashed red line and orange shading), and the no-merging subsample (dash-dotted blue line and cyan shading). Upper panel: results of the \mamp analysis. Lower panel: results of the \jei analysis. The dotted line represents orbital isotropy.} 
\label{f:comp}
\end{figure}

There is very little change in the mean values of $\rtwo$ and $\rs$ of the {\em ensemble} clusters built from these no-UG, no-merging, and high-\texttt{JClass} subsamples with respect to those obtained on the whole sample. On the other hand, there is a slight preference for the King vs. the NFW model in the best-fits to the $N(R)$ of the three subsamples, so we adopt the King model in the \mamp analysis rather than the NFW model to describe the distribution of RPS galaxies in these three subsamples (we used the NFW model instead for the whole sample, see Sect.~\ref{ss:densprof}). The core-radii of the best-fit King models are $0.56_{-0.07}^{+0.08}$ Mpc, $0.70_{-0.06}^{+0.08}$ Mpc, and $0.68_{-0.10}^{+0.11}$ Mpc for the no-UG, no-merging, and high-\texttt{JClass} subsamples, respectively.

In Fig.~\ref{f:comp} we show a comparison of the $\beta(r)$ obtained for the whole sample and the no-UG and no-merger subsamples, obtained with both the \mamp and \jei analyses. For the sake of clarity, we do not show the solution obtained for the high-\texttt{JClass} subsample as it is very similar to the other two, but with much larger confidence regions.
The $\beta(r)$ of the three subsamples are all consistent with the $\beta(r)$ of the whole sample within their 68\% confidence regions. At radii $r \lesssim 0.5 \, r_{200}$, the consistency of the four \mamp $\beta(r)$ is less good than that of the four \jei $\beta(r)$, and we attribute this to the different models used to describe the $N(R)$ in the whole sample vs. the two subsamples - in the \jei method there is no need to assume a specific model for $N(R)$.

We conclude that our result for the $\beta(r)$ of RPS galaxies is robust vs. the inclusion of merging clusters and UG, and it does not significantly depend on \texttt{JClass}.

\section{Discussion and conclusions}\label{s:disc}
We analyse a sample of 244 RPS candidates that we identify as members within the virial region of 62 nearby clusters, to determine their velocity anisotropy profile, $\beta(r)$. By using previously determined mass profiles for these 62 clusters, we stack them in projected phase-space to build an {\em ensemble} cluster. Using two methods, \mamp and \jeii, we solve the Jeans equation for the {\em ensemble} cluster and determine $\beta(r)$ for the 244 RPS cluster members that are located within $1.2 \, \rtwo$, that is within the cluster virial region, where the assumption of dynamical equilibrium is more likely to be valid.

The \mamp and \jei solutions for $\beta(r)$ of the RPS galaxies are in excellent agreement. Since they allow a good fit to the observed los velocity dispersion profile, the two velocity anisotropy profiles are acceptable dynamical equilibrium solutions. The two methods find $\beta(r) \approx 0$ near the cluster center, although the constraints are rather loose in this region due to the scarcity of RPS members near the cluster center. The orbits become more radial anisotropic with increasing cluster-centric distance, $\beta(r) \approx 0.7$ at $\rtwo$. While there have been previous suggestions that RPS galaxies move on radial orbits \citep{Solanes+01,Vulcani+17,Jaffe+18}, ours is the first direct determination of the RPS orbits. We test and confirm our results on three subsamples from which we exclude either non-validated RPS candidates (the UG), or merging clusters, or low \texttt{JClass} RPS candidates.

Previous analyses have generally found that the orbits of the general cluster population are close to isotropy near the center and become increasingly radial outside \citep[see, e.g.,][]{NK96,BK04,Benatov+06,Biviano+13,Annunziatella+16,Biviano+21}. This is particularly the case for the late-type, blue, star-foming galaxies, while early-type, red, quiescent galaxies tend instead to show more isotropic orbits, also outside the cluster center, even if not in all clusters \citep[see, e.g.,][]{BP09,MBM14,Mamon+19}. 

We compare the RPS $\beta(r)$ to those obtained by \citet{Mamon+19} for the general cluster populations in the WINGS sample of clusters, that largely overlap with the sample of clusters used in this work. We find that the $\beta(r)$ of RPS galaxies is different from those of ellipticals and S0s in WINGS clusters  and more similar to that of spirals
(see Fig.~\ref{f:betap}), but RPS galaxies have a stronger radial anisotropy than spirals at large radii - the RPS $\beta$ is almost twice as large as that of spirals at $\rtwo$ (see Fig.~\ref{f:betaprof}). 

Our results support the observational evidence by \citet{Vulcani+17} and \citet{Jaffe+18}, based on the distribution of RPS galaxies in projected phase-space, and of \citet{Salinas+23}, based on the orientation of jellyfish tails, indicating that RPS galaxies are a subset of cluster satellites on more radially elongated orbits.
Our results also support the finding of numerical simulations that radial orbits are an important requisite for ram-pressure stripping
\citep{VCBD01,DRVHB10,Tonnesen19}. In the future, we plan to compare our observed $\beta(r)$ with that of RPS galaxies from numerical simulations.

\begin{acknowledgments}
We thank the referee for her/his useful comments and suggestions.
AB acknowledges the financial contribution from the INAF mini-grant 1.05.12.04.01 {\it "The dynamics of clusters of galaxies from the projected phase-space distribution of cluster galaxies".} This project has received funding from the European Research Council (ERC) under the European Union's Horizon 2020 research and innovation program (grant agreement No. 833824). Y.J. acknowledges financial support from ANID BASAL project No. FB210003 and FONDECYT Regular No. 1230441. ACCL thanks the financial support of the National Agency for Research and Development (ANID) / Scholarship Program / DOCTORADO BECAS CHILE/2019-21190049.
\end{acknowledgments}

\bibliography{master}{}
\bibliographystyle{aasjournal}
\end{document}